\documentclass[number,12pt]{elsarticle}
\usepackage{multirow}
\usepackage{setspace, pmgraph}
\usepackage{subfigure}
\usepackage{graphicx}
\usepackage{afterpage}
\usepackage{amsmath, amsfonts, amssymb}
\usepackage{dsfont,courier}
\usepackage{verbatim}
\usepackage{longtable}
\usepackage{hyperref}
\usepackage{lineno}
\usepackage{color,soul}

\biboptions{sort}

\journal{International Journal of Fatigue}

\begin{document}

\begin{frontmatter}

\title{Multiaxial Kitagawa Analysis of A356-T6}

\author[ubc]{M. J. Roy}
\ead{majroy@interchange.ubc.ca}
\author[ENSMA]{Y. Nadot\corref{cor1}}
\ead{yves.nadot@lmpm.ensma.fr}
\author[ENSMA]{C. Nadot-Martin}
\ead{carole.nadot-martin@lmpm.ensma.fr}
\author[ENSMA]{P.-G. Bardin}
\ead{pierre-guillaume.bardin@etu.ensma.fr}
\author[ubc]{D. M. Maijer}
\ead{daan.maijer@ubc.ca}

\address[ubc]{Dept. of Materials Engineering, The University of British Columbia, Vancouver, BC, Canada V6T 1Z4}
\address[ENSMA]{Institut PPRIME - CNRS - Universit\'{e} de Poitiers - ENSMA - UPR 3346 - D\'{e}partement M\'{e}canique des Mat\'{e}riaux - T\'{e}l\'{e}port 2 - 1 Avenue Cl\'{e}ment Ader - BP 40109 - 86961 FUTUROSCOPE CHASSENEUIL Cedex - France}

\cortext[cor1]{Corresponding author, Tel. +1 604 827 5346;
Fax +1 604 822 3619}

\begin{abstract}
Experimental Kitagawa analysis has been performed on A356-T6 containing natural and artificial defects. Results are obtained with a load ratio of $R = -1$ for three different loadings: tension, torsion and combined tension-torsion. The critical defect size determined is 400 $\pm100$ $\mu$m in A356-T6 under multiaxial loading. Below this value, the microstructure governs the endurance limit mainly through Secondary Dendrite Arm Spacing (SDAS). Four theoretical approaches are used to simulate the endurance limit characterized by a Kitagawa relationship are compared: Murakami relationships [Y ~Murakami, Metal Fatigue: Effects of Small Defects and Nonmetallic Inclusions, Elsevier, 2002.], defect-crack equivalency via Linear Elastic Fracture Mechanics (LEFM), the Critical Distance Method (CDM) proposed by Susmel and Taylor [L. Susmel, D. Taylor. Eng. Fract. Mech. 75 (2008) 15.] and the gradient approach proposed by Nadot [Y. Nadot, T. ~Billaudeau. Eng. Fract. Mech. 73 (2006) 1.]. It is shown that the CDM and gradient methods are accurate; however fatigue data for three loading conditions is necessary to allow accurate identification of an endurance limit.
\end{abstract}


\end{frontmatter}
\section{Nomenclature}
\begin{description}
\item[$\sqrt{\text{area}}$] Defect size parameter defined as the square root of a defect cross-sectional area
\item[$\sqrt{\text{area}}_{\text{ref}}$] Reference defect size
\item[$b_g$] Gradient model material parameter
\item[$f_{-1}$] Endurance limit under fully reversed tension
\item[$\alpha,A,k$] Murakami parameters
\item[$t_{-1}$] Fatigue limit under fully reversed torsion
\item[$F(b/a)$] Function of defect aspect ratio ($b/a$)
\item[$H_v$] Vickers hardness
\item[$I_{1,a}$] First invariant of the stress amplitude tensor
\item[$I_{1,m}$] First invariant of the mean stress tensor
\item[$J^{\prime} _2$] Second invariant of $\underline{\underline{S_a}}$
\item[$\underline{\underline{S}}$] Deviatoric stress tensor
\item[$J_{2,\text{mean}}$] Mean value of $J^{\prime} _2$ over a period
\item[$J^{\prime}_{2,\text{max}}$] Maximum value of $J^{\prime} _2$ over a period
\item[$R$] Load ratio where $R=\sigma_{\text{min}}/\sigma_{\text{max}}$
\item[$Y$] Crack shape factor
\item[$\gamma_1,\gamma_2,\gamma_3,\beta$] Material parameters
\item[$\Sigma$] Stress tensor
\item[$\sigma_{\text{max}}$] Maximum stress over a loading cycle
\item[$\sigma_{\text{min}}$] Minimum stress over a loading cycle
\item[$\sigma_C$] Critical stress amplitude
\item[$\sigma_1$] Maximum principal stress
\item[$\sigma_2$] Minimum principal stress
\item[$\sigma_W$] Endurance limit of a specimen with a defect under tensile loading
\item[$\tau_W$] Endurance limit of a specimen with a defect under torsional loading
\item[$\sigma_{\text{eq}_{\text{Vu}}},\sigma_{\text{eq}}^\ast$] Equivalent stress
\item[$\sigma_{\text{eq}_{\text{Vu},\text{Max}}}$] Defect tip equivalent stress
\item[$\sigma_{\text{eq}_{\text{Vu},{\infty}}}$] Equivalent far-field stress
\item[$\Delta\sigma_c$] LEFM critical stress amplitude
\item[$\Delta K_{\text{th,eff}}$] Effective stress intensity factor threshold
\item[$\triangledown\sigma_{\text{eq}_{\text{Vu}}}$] Equivalent stress gradient
\item[$T$] Period length
\item[$t$] Time
\item[$\sigma_{f_e}$] Experimental endurance limit
\item[$\sigma_{f_c}$] Calculated endurance limit
\item[$\delta_c$] Percent difference between calculated and experimental endurance limits such that $\delta_c=(\sigma_{f_e}-\sigma_{f_c})/\sigma_{f_e}\times 100$\%
\end{description}

\section{Introduction}\label{sec:intro}
The tensile fatigue behaviour of A356-T6 has been the subject of study by a number of researchers \cite{Buffiere.01,Ludwig.03,Fan.03,Gall.00,Brochu.10,Chan.03,Zhu.07}. In almost all related studies, casting defects such as intermetallic inclusions, porosity, shrinkages and oxide films have been shown to be present at the origin of the failure. In cases where defects were not present, the basic microstructure \cite{Gao.04,McDowell.03} has been shown to determine the fatigue life. The presiding microstructural factor in this latter case is Secondary Dendrite Arm Spacing (SDAS) which decides the overall strength of the material. By processing samples of A356 via Hot Isostatic Pressing (HIP) to eliminate porosity, Gao et al. \cite{Gao.04} compared the influence of SDAS and porosity on the tensile endurance limit. It was found that the fatigue limit was significantly increased either through HIP or halving the SDAS. While the effects of porosity and SDAS are not mutually exclusive, the role SDAS plays is less important when the material contains defects. Another microstructural characteristic that participates in fatigue mechanisms is the secondary eutectic phase. Fatigue crack initiation has been found to occur at silicon particles within this phase \cite{Brochu.10,McDowell.03} in samples that were free of defects. However, in the majority of prior studies, the authors did not explicitly quantify the critical defect size. One exception is the work of Brochu et al. \cite{Brochu.10} where an experimental Kitagawa relationship was developed for rheocast A357, a similar alloy to A356. This study demonstrated a critical defect size of 150 $\mu$m under fully reversed ($R=-1$) tensile loading. In terms of multiaxial fatigue behaviour, only two previous studies are available for A356 \cite{McDowell.03,De-Feng.08}. De-Feng et al. \cite{De-Feng.08} performed tension-torsion fatigue testing on thin-wall specimens in the low-cycle regime ($10^4$ cycles). McDowell et al. \cite{McDowell.03} performed torsional High-Cycle Fatigue (HCF) testing, however these tests were conducted with deformation control.

The fatigue life prediction of nominally defective materials such as A356 is of great importance to industry and has been the subject of considerable study. Linear Elastic Fracture Mechanics (LEFM) has been shown to only apply to the study of long cracks where stress fields are homogeneous and not affected by local plasticity, making it inappropriate to be applied on short cracks \cite{Susmel.04}. Therefore, treating defects as cracks and leveraging analytical approaches based on LEFM may not always be appropriate. Using experimental assessments of local plasticity through microhardness measurements, Murakami  \cite{Murakami.02} has shown that fatigue behavior can be assessed for some materials through empirical relations based on applied stress and hardness. In terms of more advanced computational and analytical assessments, there are two stress-based approaches as highlighted by Atzori et al. \cite{Atzori.05}. The first are local methods which dictate that an effective stress must be reached around a defect to affect the fatigue resilience \cite{Lazzarin.97,Taylor.99} or nominal stress methods where the fatigue or endurance limit of the material is defined in terms of a nominal applied stress \cite{Atzori.00}. Most recently, Critical Distance Theory \cite{Susmel.08} has been shown to successfully correlate fatigue behavior with a critical distance from a defect that is material dependent. Other studies have shown that the stress gradient around a defect is a more practical indicator of fatigue resilience \cite{Papadopoulos.96,Nadot.06,Gadouini.08}.

The objective of the present study is to investigate the influence of casting defects on the HCF behaviour of A356-T6. Kitagawa-type analysis is performed with experimental results for three different scenarios: tension, torsion and combined loading. The critical defect size from a multiaxial standpoint is then defined. Finally, four different approaches to simulating the evolution of the endurance limit with increasing defect sizes are compared for each of the loading cases. This is done to determine the best method for simulating the endurance limit based on defect size for this industrially relevant alloy.
\section{Material and experimental conditions}
The material employed in this study was Low-Pressure Die Cast (LPDC), strontium modified A356 (Al-7Si-0.3Mg) in the T6 condition with a typical chemical composition given in Table \ref{table:composition}. The majority of fatigue specimens were cut from a wedge-shaped casting, and a lesser number were cut directly from an automotive wheel casting. While both casting types were made with permanent steel dies, the wheel casting was actively cooled during solidification while the wedge casting was  left to cool passively. As these two casting types encompass a wide range of solidification conditions, so too did the specimens from a defect and microstructural standpoint. The fatigue behaviour characterized in the current work is thus expected to represent material produced from a range of industrial practices.
\begin{table}[h!]
\centering
\caption{A356 composition in wt-\%\label{table:composition}}
    \begin{singlespacing}
    \begin{tabular}{lllll}
    \hline
    \textbf{Element} & Si & Mg & Na & Sr\\
    \textbf{Range (wt-\%)} & 6.5-7.5 & 0.25-0.4 & $\sim$0.002 & $\sim$0.005\\
    \hline
\end{tabular}
\end{singlespacing}
\end{table}
\subsection{Material preparation and microstructure}\label{sec:prepmatl}
\begin{figure}
\centering
      \subfigure[Wedge mold and the resultant wedge casting\label{fig:mold}]{\includegraphics[width=2.5 in]{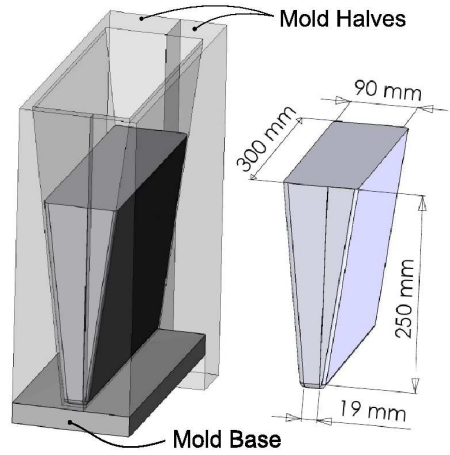}}\\
      \hspace{0.01 in}
      \subfigure[The three different types of fatigue specimens and their respective geometries: (i) tension-torsion, (ii) torsion and (iii) tension type.\label{fig:specimens}]{\includegraphics[width=2.5 in]{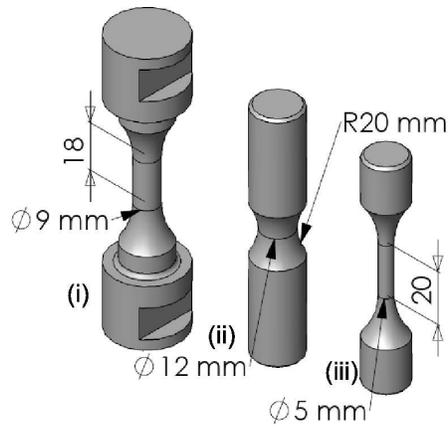}}
      \caption[]{Wedge mold, resultant casting and the different types of fatigue specimens used\label{fig:mold_specimens}}
\end{figure}

A wedge was generated via the steel die, shown in Figure \ref{fig:mold}. The casting produced was 250 mm in height, 300 mm in length and had a 7$^{\circ}$ wedge angle. The motivation for this geometry was to create a gradient in cooling rate varying with height in the wedge. A more refined microstructure was expected at the base of the wedge where the cooling rate was the highest as compared to a coarser microstruture at the top where the cooling rate was the lowest. One half of the wedge was devoted to tension-torsion fatigue specimens only, and the other half had a mixture of tension, torsion, and combined tension-torsion fatigue specimens (Figure \ref{fig:specimens}) drawn from it. All specimens were then heat treated to a T6 condition after being removed from the wedge block with the following schedule: solutionized at 538$^{\circ}$C for 3 hours, quenched in water at 60$^{\circ}$C, and finally artificially aged at 150$^{\circ}$C for 3 hours. Tension-torsion specimens were extracted from the spokes of the commercially cast automotive wheel which had the same material composition as the wedge casting. The wheel was subjected to the same T6 treatment before specimens were extracted. Tensile testing of wheel material resulted in a modulus of elasticity of 66 GPa, a yield strength of 164 MPa and an ultimate tensile strength of 317 MPa.

Hemispherical artificial defects were introduced to the middle of the gage sections of six specimens via Electro-Discharge Machining (EDM). In total, four tensile and two torsion specimens were drawn from the bottom of the wedge and had artificial defects applied post heat-treatment. This technique of generating artificial defects has been qualified in other crack propagation investigations \cite{Billaudeau.environmental.04,Nadot.06}, and the size of these defects is presented in Table \ref{table:fatiguespecimens}. A cross-section of a typical artificial defect is given in Figure \ref{fig:A1}.

Metallographic specimens were cut from the wedge and wheel at locations corresponding to where fatigue specimens were taken. This resulted in samples from 8 evenly spaced locations through the height of the wedge, and 3 locations corresponding to the wheel specimens. On each of these samples, the SDAS  and porosity were characterized via optical microscopy and Clemex Vision PE software. Porosity was assessed by both the area percent and the $\sqrt{\text{area}}$ of each pore. Specimens and metallography results were grouped to provide four families of specimens: one for the wheel, and one each for the low, mid-range and high SDAS wedge specimens (Table \ref{table:SDAS_pores}). The average macrohardness value $H_v$ value was found for each specimen family with a load of 5 kg and a minimum of 30 indents on two specimens per specimen family. The average $H_v$ value for all material was 85 MPa. Both the $\sqrt{\text{area}}$ parameter of all defects and the overall hardness of this material are within the range as specified by Murakami \cite{Murakami.02}.

The secondary dendrite arm spacing showed a definite increase from the bottom of the wedge to the top coinciding with the cooling rate differential imposed by the casting practice. Average porosity remained relatively uniform throughout the wedge based on the percent area, with the peak pore size found through metallographic analysis was 94 $\mu$m.  These porosity measurements show that the middle of the wedge had higher porosity than the top and bottom of the wedge as characterized by the area percent and maximum pore size. Furthermore, while the bottom of the wedge approached the same SDAS as the wheel, the porosity in the wedge was approximately double that of the wheel casting.

\begin{table}[h!]
\centering
    \begin{singlespacing}
    \begin{tabular}{llccc}
    \hline
    \multirow{3}{*}{Family} & SDAS & \multicolumn{2}{c}{Porosity} &$H_v$ $\pm 1$ St. Dev.\\
                            & ($\mu$m)                               & Area  & Max. $\sqrt{\text{area}}$ &(MPa)\\
                            &                                       &(\%) & ($\mu$m) &\\
    \hline
Wheel (W)	&37 $\pm 8$& 0.0603&52& 86.15 $\pm2$\\
Wedge Bottom (A/B)	&42  $\pm$11& 0.1237&32&78.9 $\pm5$\\
Wedge Middle (M)	&   59   $\pm$21 & 0.1244&94&83.1 $\pm3$\\
Wedge Top (T)	&68   $\pm$26& 0.1232&42&89 $\pm7$\\
\hline
\end{tabular}
\end{singlespacing}
\caption[]{Secondary dendrite arm spacing, porosity measurements and Vickers hardness values for all material. Family A denotes specimens extracted from the bottom of the wedge which had hemispherical artificial defects applied via EDM. Family B denotes specimens extracted from the bottom of the wedge containing only natural defects.\label{table:SDAS_pores}}
\end{table}

\subsection{Fatigue testing}
All fatigue tests were run in load controlled mode with a sinusoidal signal and fully reversed ($R = -1$). The combined tension-torsion specimens were tested on an Instron servo-hydraulic test platform at 11 Hz, while the pure torsion and pure tension specimens were tested at 45 Hz on a Amsler-Vibraphore machine.
\begin{table}[h!]\label{table:fatiguespecimens}
\centering
\caption{Test history for all fatigue specimens. Specimens A1 through A6 contained hemispherical defects applied via EDM.\label{table:fatiguespecimens}}
    \begin{singlespacing}
    \begin{tabular}{ccccllcl}
    \hline
    \multicolumn{2}{c}{Specimen}  & \multicolumn{2}{c}{Loading at Failure (MPa)} & \multicolumn{2}{c}{Step} & $N_f$ & $\sqrt{\text{area}}$\\
    Name & Type$^{\text{a}}$                   &            $\sigma_a$ & $\tau_a$ & Number & MPa/step & ($\times 10^5$) &($\mu$m)\\
                                \hline
    W1 & TT & 0 & 90 & 3 & 5 & 7.22&59$^{\text{c}}$\\
    W2 & TT & 0 & 85 & 1$^{\text{b}}$ & N/A & 3.00&59$^{\text{c}}$\\
    W3 & TT& 70 & 70 & 2 & 5& 1.05&59$^{\text{c}}$\\
    \hline
    B1 & TT& 70 & 70 & 2 & 5& 0.76&90$^{\text{c}}$\\
    B2 & To& 0 & 70 & 2 & 5& 3.98&39$^{\text{c}}$\\
    B3 & TT& 0 & 100 & 3 & 10& 1.51&30$^{\text{c}}$\\
    B4 & TT& 0 & 110 & 2 & 10& 8.83&38$^{\text{c}}$\\
    \hline
    M1 & TT& 95 & 0 & 3 & 5& 0.79&90$^{\text{c}}$\\
    M2 & TT& 65 & 65 & 2 & 5& 2.46&514\\
    M3 & TT& 70 & 70 & 3 & 5& 5.26&53$^{\text{c}}$\\
    M4 & To& 0 & 60 & 2 & 10 & 3.25&531\\
    M5 & To& 0 & 55 &1$^{\text{b}}$ & N/A & 2.27&90$^{\text{c}}$\\
\hline
    T1 & TT& 65 & 65 & 5 & 5& 4.24&112$^{\text{c}}$\\
    T2 & TT& 65 & 65 & 2 & 5& 1.29&265\\
    T3 & TT& 65 & 65 & 1 & N/A& 9.08&300\\
    T4 & TT& 60 & 60 &1$^{\text{b}}$ & N/A&4.05&496\\
    T5 & To& 0 & 50 & 2 & 10& 4.84&265\\
    T6 & Te& 90 & 0 & 5 & 10& 6.63&372\\
    T7 & To& 0 & 50 &1$^{\text{b}}$ & N/A&7.33&310\\
\hline
    A1 & Te& 90 & 0 & 1$^{\text{b}}$ & N/A& 4.24&398\\
    A2 & Te& 90 & 0 & 3 & 10& 1.29&514\\
    A3 & Te& 80 & 0 & 4 & 10& 9.08&740\\
    A4 & Te& 70 & 0 & 2 & 10& 5.26&760\\
    A5 & To& 0 & 70 & 4 & 10& 4.84&465\\
    A6 & To& 0 & 50 & 2 & 10& 6.63&708\\
\hline
\end{tabular}
\end{singlespacing}
\newline
$^{\text{a}}$\small{TT: tension-torsion, To: torsion, Te: tension}\\
$^{\text{b}}$\small{Failure before $10^6$ cycles}\\
$^{\text{c}}$\small{Estimated based on the maximum $\sqrt{{\text{area}}}$ found with metallography}
\end{table}
atigue testing was conducted by employing the step technique as originally outlined by Maxwell and Nicholas \cite{Maxwell.99}. Each specimen underwent cyclic loading starting at a given stress amplitude assumed to be below the expected endurance limit, and samples that did not fail after $10^6$ cycles were cycled again at a stress amplitude one step higher. This was repeated until the specimen ruptured after less than $10^6$ cycles. The endurance limit ($\sigma_{f_e}$) was taken to be the applied stress amplitude ($\sigma_a$, $\tau_a$) of the final step.

For example, specimen W1 underwent $10^6$ cycles at $\tau_a=$80 MPa without failure. The stress amplitude was increased by 5 MPa to $\tau_a=$85 MPa, where it also withstood $10^6$ cycles. The stress amplitude was increased again by 5 MPa to $\tau_a=$90 MPa and the sample failed after 722,000 cycles, having withstood 3 loading steps of 5 MPa per step. Table 3 describes the details of the step testing testing conducted. While the results obtained from this testing are not endurance limits from a statistical standpoint, step testing is the only technique that permits the evaluation of an `endurance limit' for a natural defect of unknown size. Furthermore, since the material is non-ferrous and there were specimens with similar microstructure requiring more steps than others to rupture at the same stress amplitude, the effect of coaxing is considered negligible for this material. In the current work, the term `endurance limit' is defined as the stress level at fracture for one million cycles.
\section{Experimental results}\label{sec:ExpKit}
Experimental fatigue test results are given in Figures \ref{fig:TensKit}, \ref{fig:TorKit} and \ref{fig:TenTorKit} in the form of bi-linear Kitagawa diagrams for each of the loading cases. The fracture surfaces were first examined macroscopically, and these observations guided Scanning Electron Microscopy (SEM) analysis of the initiation area. The goal of SEM observations (Figures \ref{fig:TensSEM}, \ref{fig:TorSEM} and \ref{fig:TenTorSEM}) was to clearly identify the initiation site and where possible, measure the initiating defect. Multiaxial fracture surfaces were found to be tortuous and indentification of the defect at the origin of the fatal crack was non-trivial. The systematic methodology employed to find the defect and correctly identify the area of the defect on the Kitagawa diagrams presented is as follows:
\begin{itemize}
\item	Optical observation of the fracture surface employed to determine both the fatigue and final failure zones.
\item	The fatigue zone of the fracture surface was found to be clear enough to identify a given initiation area. The `river marks' observed on the surface converged to this unique zone.
\item	If there was a clearly defined defect, then the size was assessed using the $\sqrt{\text{area}}$ parameter directly from the fracture surface. This was done regardless of the position of the defect relative to the free surface.
\end{itemize}
For specimens where the initiation site was unidentifiable, the initiating feature was estimated as the maximum $\sqrt{\text{area}}$ of porosity found via metallography (Section \ref{sec:prepmatl}). The use of these results in the Kitigawa analysis is thus speculative. For specimens where there were multiple initiation sites, the largest identifiable defect closest to the surface was characterized. The results of this analysis for all specimens are summarized in the final column of Table \ref{table:fatiguespecimens}.
\subsection{Tensile results}\label{sec:TensResults}
\begin{figure}
\centering
      \includegraphics[width=4 in]{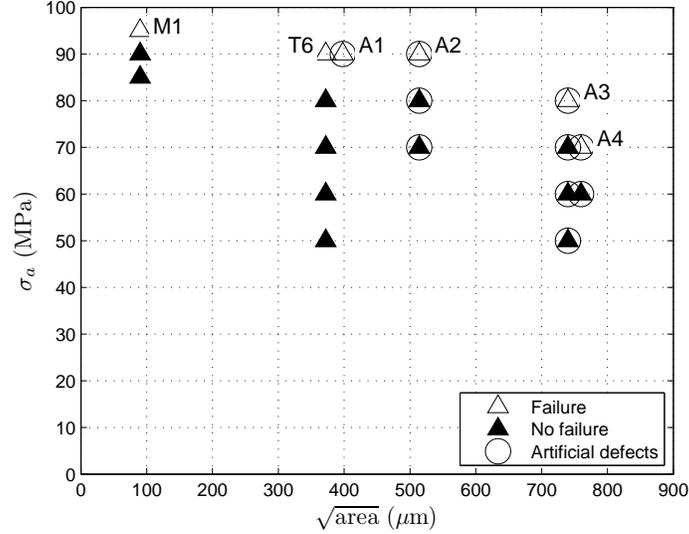}
      \caption[]{Kitagawa diagram for A356-T6 under tension. Specimens A1 through A4 had artificial defects applied. \label{fig:TensKit}}
\end{figure}

\begin{figure}
\centering
      \subfigure[\label{fig:M1}]{\includegraphics[width=2.4 in]{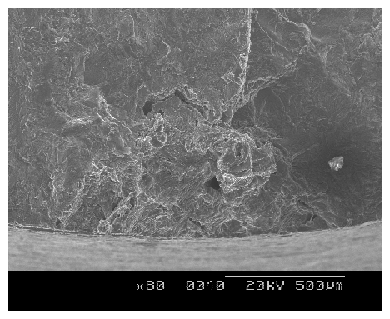}}
      \hspace{0.01 in}
      \subfigure[\label{fig:T6}]{\includegraphics[width=2.4 in]{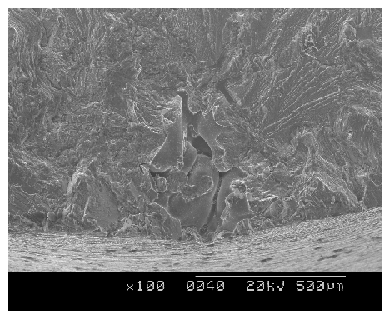}}\\
      \vspace{0.01 in}
      \subfigure[\label{fig:A1}]{\includegraphics[width=2.4 in]{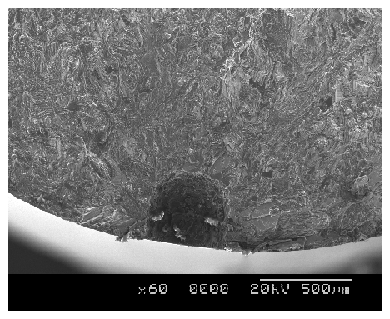}}\\
      \caption[]{SEM observations of initiation sites on tensile specimens (a) M1 and (b) T6 having natural defects; (c) A1 having an artificial defect\label{fig:TensSEM}}
\end{figure}

Figure \ref{fig:TensKit} presents the experimental Kitagawa relationship under pure tension. In all samples, the initial defect size was readily identifiable on the fracture plane that was found to be perpendicular to the direction of the maximum principal stress. The primary finding from the tensile Kitagawa curve is that the critical defect size is relatively large: specimens T6, A1 and A2 have a very low impact on the endurance limit (8 \% reduction). These tensile specimens show that the material appears to be sensitive to defects only when $\sqrt{\text{area}}$ is greater than than 500 $\mu$m.
Figure \ref{fig:TensSEM} presents the fracture surfaces for specimens M1, T6 and A1. Specimen M1 shows a oxide-related defect at the origin of the crack (Figure \ref{fig:M1}) linking to subsurface porosity. Specimen T6 failed due to a 400 $\mu$m shrinkage pore that intersected with the surface of the sample (Figure \ref{fig:T6}), while specimen A1 failed due to a 400 $\mu$m artificial defect (Figure \ref{fig:A1}). Since specimens T6 and A1 demonstrated the same endurance limit, it is concluded that the area parameter is able to correlate different types of defects, independent of the nature of the defect. Furthermore, as these two specimens exhibit the same endurance limit despite being at the upper and lower end of the range of SDAS, this defect size is used as the reference defect size ($\sqrt{\text{area}}_{\text{ref}}$). Nevertheless, this finding should be verified with larger defects having a greater impact on the endurance limit. In terms of artificial defects, fracture surfaces for specimens A2, A3 and A4 were very similar to A1, showing that the artificial defect was inarguably the initiation point.
\subsection{Torsion results}\label{sec:TorsionResults}
\begin{figure}
\centering
      \includegraphics[width=4 in]{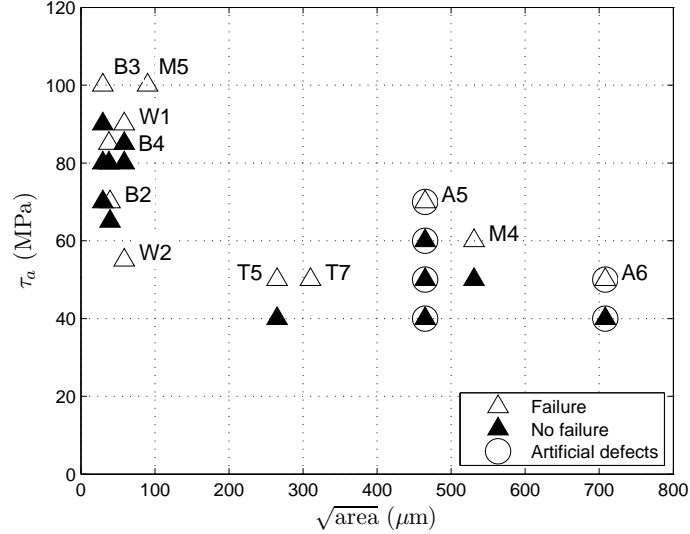}
      \caption[]{Kitagawa diagram for A356-T6 under torsion. Specimens A5 and A6 had artificial defects applied.\label{fig:TorKit}}
\end{figure}

\begin{figure}
\centering
      \subfigure[\label{fig:W2macro}]{\includegraphics[height=1.75 in]{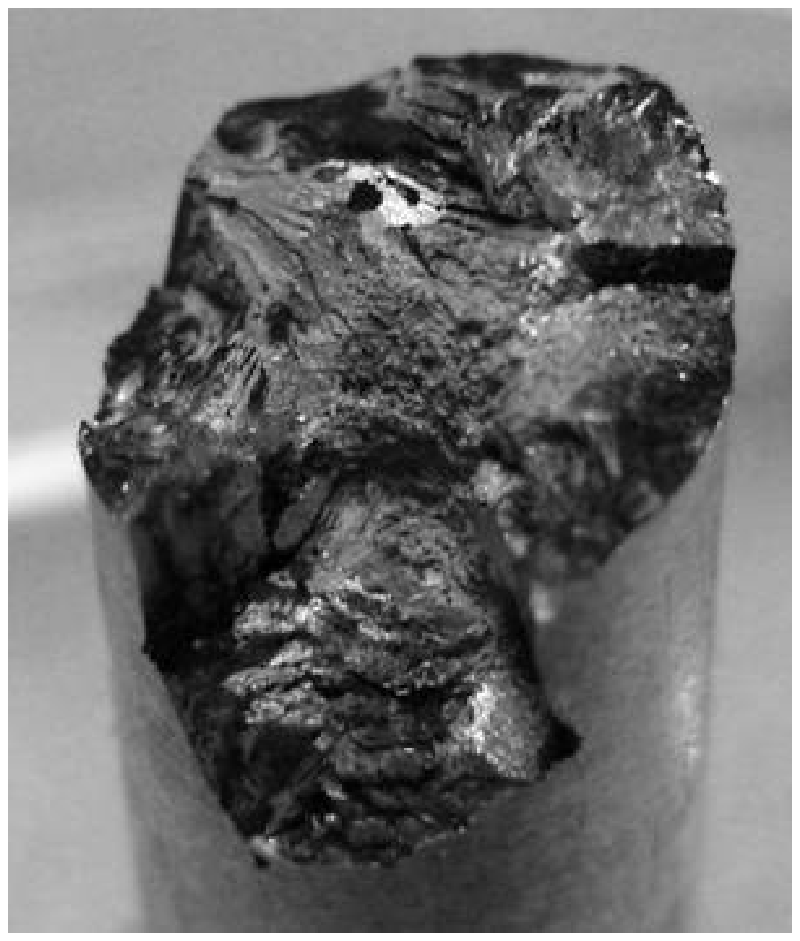}}
      \hspace{0.01 in}
      \subfigure[\label{fig:W2init}]{\includegraphics[height=1.75 in]{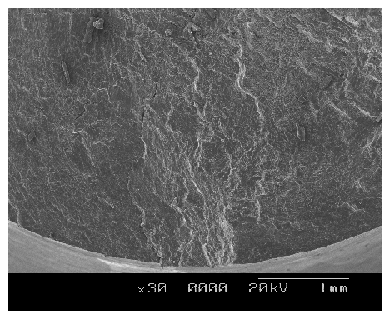}}\\
      \vspace{0.01 in}
      \subfigure[\label{fig:T7init1}]{\includegraphics[height=1.75 in]{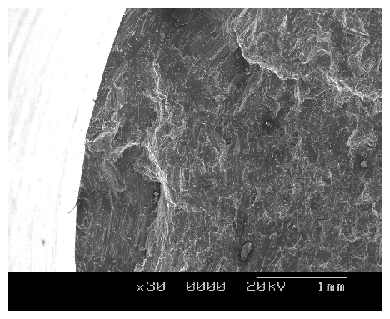}}
      \hspace{0.01 in}
      \subfigure[\label{fig:T7init2}]{\includegraphics[height=1.75 in]{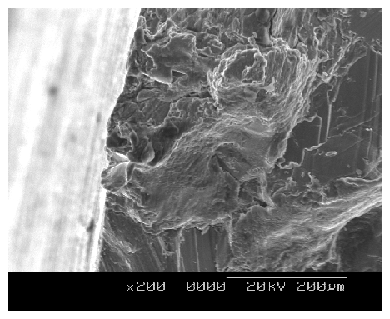}}
      \vspace{0.01 in}
      \subfigure[\label{fig:M4}]{\includegraphics[height=1.75 in]{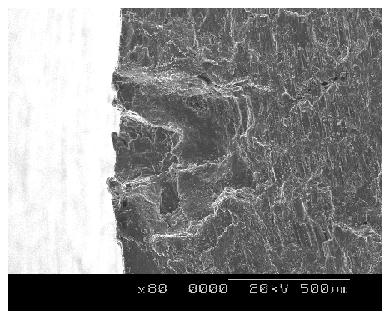}}
      \caption[]{Observations of initiation site defects on torsion specimens (a,b) W2 (c,d) T7 and (f) M4\label{fig:TorSEM}}
\end{figure}

The Kitagawa diagram for torsion testing, shown in Figure \ref{fig:TorKit}, is similar to the tensile results (Figure \ref{fig:TensKit}). The experimental points presented on the curve below 100 $\mu$m are for specimens that were unsuccessfully classified by fractography. The samples have been separated along the horizontal axis based on the porosity assessment of their location in the wedge to render individual tests identifiable from a stress amplitude standpoint. As the endurance limits vary from 55 to 95 MPa for specimens with unidentifiable defects, the Kitagawa diagram under torsion exhibits a large amount of scatter as compared to the tensile results.
The main complication in identifying defects in the sub 100 $\mu$m range is the tortuous nature of the fracture surface, as shown in Figure \ref{fig:W2macro}. Cracking was activated on two planes of maximum shear such that the final fracture surface reveals multiple initiation sites. While multiple initiation sites could explain the scatter seen in the Kitagawa plot, careful examination of the fracture surface at each suspected initiation point (Figure \ref{fig:W2init}) did not always result in an identifiable defect. An attempt to link the presence of porosity to the multiple initiation sites was made with specimens B2, M5, T5 and T7 by metallography performed on sectioned fracture surfaces. There was little to no deviation found from the porosity measurements given in Table \ref{table:SDAS_pores}.  In light of these findings, the critical defect size is difficult to assess. Specimens with identifiable defects show a definite decrease in endurance limit beyond 300 $\mu$m, which is smaller than under tension. As is the case with the tensile testing, this critical defect size should be clarified by other tests on samples containing 300 $\mu$m or larger artificial defects.

Figure \ref{fig:TorSEM} presents an example fracture surface for a specimen under pure torsion and gives an indication of the typical challenge experienced in determining the defect size. It is surmised that the crack lips contact during crack propagation, and an excess of oxide fouls the fracture surface masking both the defect size and type. Macroscopically speaking, this black oxide appears throughout the gauge section of the pure torsion specimens. Figure \ref{fig:W2macro} shows the fracture surface of specimen W2 and Figure \ref{fig:W2init} is an image of a crack origination point on the same specimen which is absent of any identifiable defect. Specimen T7 showed multiple initiation points, but no defects were observed at the crack origins (Figure \ref{fig:T7init1}). The defect attributed to specimen T7 (Figure \ref{fig:T7init2}) was an identifiable defect located on the fracture plane of the main crack. Specimen M4 provides an example of a large, readily identifiable defect at the origin of failure (Figure \ref{fig:M4}).
\subsection{Combined tension-torsion results}
\begin{figure}
\centering
      \includegraphics[width=4 in]{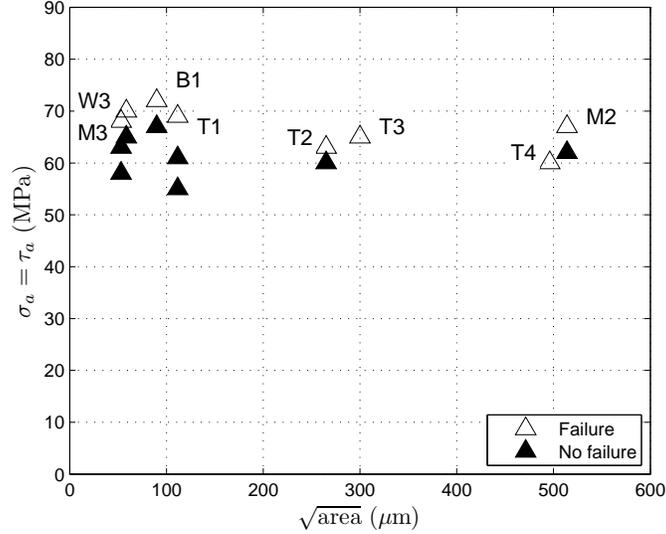}
      \caption[]{Kitagawa diagram for A356-T6 under tension-torsion\label{fig:TenTorKit}}
\end{figure}

\begin{figure}
\centering
      \subfigure[\label{fig:B1}]{\includegraphics[height=1.75 in]{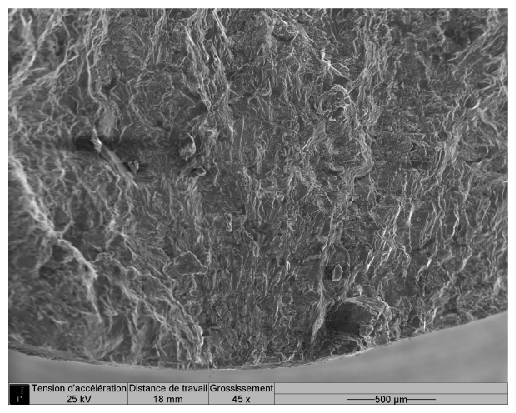}}
      \hspace{0.01 in}
      \subfigure[\label{fig:M2}]{\includegraphics[height=1.75 in]{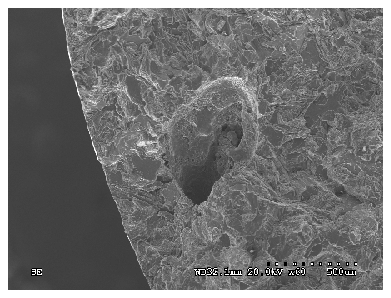}}\\
      \vspace{0.01 in}
      \subfigure[\label{fig:T3}]{\includegraphics[height=1.75 in]{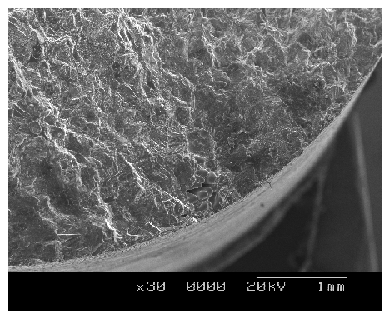}}\\
      \caption[]{SEM observation of initiation sites on tension-torsion specimens (a) B1 (b) M2 (c) T3\label{fig:TenTorSEM}}
\end{figure}

The combined tension-torsion Kitagawa diagram, presented in Figure \ref{fig:TenTorKit}, includes results for specimens with natural defects only. The macroscopic fracture surfaces were similar to those of the tensile specimens: a flat surface in the plane perpendicular to the direction of the maximum principal stress with clear, identifiable initiation sites the exception of specimen B1 (Figure \ref{fig:B1}). For the specimens tested, the Kitagawa diagram shows a very small influence of a 500 $\mu$m defect such as specimen M2 (Figure \ref{fig:M2}) exhibiting a large 500 $\mu$m subsurface pore. Below this size, there is no apparent influence of defects on the endurance limit.
Figure \ref{fig:T3} reveals a shrinkage void at the origin of the failure on sample T3. It is of interest to highlight that specimen M2 has approximately the same endurance limit (~67 MPa) as specimen T3, reinforcing the independence of defect type and dependence of defect size characterized by $\sqrt{\text{area}}$ on the overall fatigue life.
\subsection{Experimental summary}\label{sec:ExpSum}
The experimental results demonstrate an average endurance limit of 90 MPa for pure tension ($f_{-1}$) and 80 MPa for pure torsion ($t_{-1}$). Furthermore, these results clearly indicate that the defect size necessary to affect the endurance limit is relatively large. The critical defect size ranges from 300 $\mu$m for pure torsion to 500 $\mu$m for pure tension and combined tension-torsion. This suggests such that an overall critical defect size necessary to diminish the multiaxial endurance limit is $\sqrt{\text{area}}_{\text{ref}}=400$ $\mu$m. Together with this reference defect size, the tension and torsion endurance limits will be used in identifying model parameters in the following sections.

Another important qualitative observation is that when the defect is characterized by the $\sqrt{\text{area}}$ parameter, an artificial defect diminishes the endurance limit in the same manner as naturally occurring defects. Therefore, the $\sqrt{\text{area}}$ is a powerful method to describe the size influence of a defect. However, analyzing the complicated topology of the fracture surfaces coupled with multiple initiation sites precludes an accurate assessment of the size of critical defects.

Within a Kitigawa framework, the pure torsion results show a great deal of scatter, and indicate the need for further investigation. In light of the difficulty in assessing critical defect size for some specimens, more experimentation is necessary to categorically determine the effects of small defects on the endurance limit of this material. In terms of large defects, as the combined tension-torsion results do not clearly show a drop in endurance limit as compared to the tension results, further assessment of this loading scenario and material is also desirable.
\section{Simulation of multiaxial Kitagawa relationships}
Four standard models to predict the endurance limit of defective materials were employed to simulate the behaviour of A356-T6. What follows is a presentation of each model, followed by a comparison of the results of each model to experimental data in Section \ref{sec:comparison}. The following models are able to predict the Kitagawa relationships for tension, torsion and combined loading:
\begin{enumerate}[1.]
\item Linear Elastic Fracture Mechanics (LEFM) approach whereby a defect is considered equivalent to a crack;
\item Murakami relationships \cite{Murakami.02} employing an empirical function of the Vickers hardness, load ratio and defect size;
\item Critical Distance Method (CDM) based on the stress field around a defect proposed by Susmel and Taylor \cite{Susmel.08}; and
\item Gradient approach based on the stress gradient surrounding the defect proposed by Nadot \cite{Nadot.06,Gadouini.08}
\end{enumerate}
In the case of the CDM and gradient approaches, a multiaxial fatigue criterion and a description of the stress distribution around a defect is necessary. For the multiaxial fatigue criterion, the equivalent stress criterion recently proposed by Vu \cite{Vu.2010} will be used. This criterion determines the multiaxial behaviour for complex loadings using an invariant approach. The equivalent stress criterion as proposed by Vu is generalized to account for phase-shifted loading effects for both hard and soft metallic materials. This equivalent stress is given as:
\begin{equation}\label{eq:VuEquivalentStress}
\sigma_{\text{eq}_{\text{Vu}}}=\sqrt{\gamma_1 J^{\prime}_{2}(t)^2+\gamma_2 J_{2,\text{mean}}^2 + \gamma_3 I_f(I_{1,a},I_{1,m})} \leq \beta
\end{equation}
where $J^{\prime}_{2}(t)$ invokes the stress amplitude tensor deviator:
\begin{equation}
J^{\prime}_{2}(t)=\sqrt{\frac{1}{2}\underline{\underline{S}}(t):\underline{\underline{S}}}(t)=\sqrt{\frac{\Sigma_{xx}(t)^2}{3}+\Sigma_{xy}(t)^2}
\end{equation}
For fully reversed loading, $J^{\prime}_{2}(t)=J^{\prime}_{2_\text{max}}$.
While $J_{2,\text{mean}}$ is evaluated over a full period $T$:
\begin{equation}
J_{2,\text{mean}} = \frac{1}{T}\int\limits_{0}^T J^{\prime}_2 (t)dt
\end{equation}
Without mean stress, $J_{2,\text{mean}}=0$. The first invariant function, $I_f(I_{1,a},I_{1,m})$, is a linear function of the first invariants of the stress amplitude and mean stress where
\begin{gather}
I_f(I_{1,a},I_{1,m})= I_{1,a} + \alpha I_{1,m}\\
I_{1,a}=\frac{1}{2}\lbrace\max_{\substack{t\in T}}I_1(t)-\min_{\substack{t\in T}}I_1(t)\rbrace\\
I_{1,m}=\frac{1}{2}\lbrace\max_{\substack{t\in T}}I_1(t)+\min_{\substack{t\in T}}I_1(t)\rbrace
\end{gather}
For fully reversed loading conditions, $I_{1,m}=0$ and $I_f=I_{1,a}=\Sigma_{xx}+\Sigma_{yy}+\Sigma_{zz}$.
The values of $\beta$ and $\gamma_{1-3}$ are material properties based on the strength of the material and were empirically identified by Vu with the torsional endurance limit $t_{-1}$, the tensile endurance limit $f_{-1}$ and the ultimate tensile strength. As the ultimate tensile strength of A356-T6 is less than 750 MPa, $\gamma_1=0.65$. The torsional endurance limit at $10^6$ cycles for A356-T6 is taken to be 80 MPa, therefore $t_{-1}=\beta=80$ MPa. The final material coefficient, $\gamma_3$ is given as:
\begin{equation}
\gamma_3=\frac{(t_{-1})^2-\dfrac{(f_{-1})^2}{3}}{f_{-1}}
\end{equation}

The second input data required by both the CDM and gradient approaches is the stress distribution around the defect. To aid in this calculation, the defect is considered a spherical void in an infinite, homogeneous, isotropic, elastic matrix subjected to uniform stress at infinity. Local stresses around the void are calculated using the equivalent inclusion method \cite{Eshelby.57} and solved analytically \cite{Mura.87}. This analytical description of the elastic stress distribution around a void shows excellent correlation with that described by the Finite Element Method.
\subsection{LEFM approach}
LEFM describes a crack propagation threshold through the amplitude of the stress intensity factor, which is a function of crack length and stress amplitude $\Delta \sigma_c$. The defect size characterized by the area parameter is transformed to an equivalent semi-circular crack. The relationship between the endurance limit and defect size is therefore given by:
\begin{equation}
\Delta \sigma_c = \frac{\Delta K_{\text{th,eff}}}{Y\sqrt{{\sqrt{2 \pi}\sqrt{\text{area}}}}}
\end{equation}
where $Y$ is the crack shape factor and $\Delta K_{\text{th,eff}}$ is the effective stress intensity factor threshold for crack propagation. The values used for A356-T6 were $Y=2/\pi$ and $\Delta K_{\text{th,eff}}=1.5$ MPa$\sqrt{\text{m}}$. These were not experimentally determined, but were synthesized from a large compilation of published data \cite{Buffiere.01,Couper.90,Atzori.04,Dabayeh.96,Zhu.07,Kumar.10}. While the LEFM approach requires an experimentally determined effective threshold stress intensity factor and a description of the defect shape to estimate the crack shape factor, this approach can be used to account for multiaxial complex loading and account for load ratio effects.
\subsection{Murakami relationships}\label{sec:MurakamiDesc}
Murakami \cite{Murakami.02} proposed the use of the $\sqrt{\text{area}}$ parameter to describe the size of a surface defect. The basis for this representation were non-propagating crack observations within a small stressed region surrounding a defect. From this study, the endurance threshold supposedly corresponds to the crack growth threshold. It was also demonstrated that the maximum stress intensity factor $K_{I_{\text{max}}}$ is linearly related to the $\sqrt{\text{area}}$ parameter for different crack geometry. Furthermore, it was found that the fatigue crack growth threshold could also be correlated to a given value of Vickers hardness. The result is an empirical equation based on the defect size ($\sqrt{\text{area}}$) and macrohardness to predict the endurance limit of materials containing small defects:
\begin{equation}
\sigma_W=\frac{A\left(H_v+120\right)}{\left(\sqrt{\text{area}}\right)^{1/6}}\left(\frac{1-R}{2}\right)^{\alpha}
\end{equation}
where $\alpha=0.226+H_v \times 10^{-4}$ and $A=1.43$ or 1.56 for surface and internal defects, respectively. For torsional loading and surface defects, the endurance limit is expressed as:
\begin{equation}
\tau_W=\frac{0.93\left(H_v+120\right)}{F(b/a)\left(\sqrt{\text{area}}\right)^{1/6}}\left(\frac{1-R}{2}\right)^{\alpha}
\end{equation}
where $F(b/a)=0.8397$ for spherical defects. For combined tension-torsion loading, the endurance limit is given by the following:
\begin{equation}
\sigma_1+k\sigma_2=\frac{A\left(H_v+120\right)}{\left(\sqrt{\text{area}}\right)^{1/6}}\left(\frac{1-R}{2}\right)^{\alpha}
\end{equation}
with $k=-0.18$ for cracks emanating from a round defect. There are two parameters that are required for the preceding relationships: the macroscopic Vickers hardness (85 MPa) as well as the value of $F(b/a)$, which is a constant for spherical defects. While requiring minimal parameters to be identified, the work of Murakami is limited due to the inability of the stress state description to allow for general multiaxial loading.
\subsection{Critical Distance Method}\label{sec:critdistance}
Based on the methodology proposed by Susmel and Taylor \cite{Susmel.08}, the main tenet of this approach is to describe the influence of the defect through computation of the equivalent fatigue stress at a given distance from a defect. This requires an evaluation of the stress field surrounding a defect as defined previously (Eq. \ref{eq:VuEquivalentStress}). As the equivalent stress is calculated at the critical distance $L/2$ from the tip of the defect, the criterion is expressed as:
\begin{equation}
\max_{\sigma=\sigma_C}\left(\sigma_{\text{eq}_{\text{Vu}}}\left(\frac{L}{2}\right)\right)=\beta
\end{equation}
where $\sigma$ is the nominal applied stress. The critical distance $L/2$ is therefore the point where $\sigma_{\text{eq}_{\text{Vu}}}=\beta$. In order to identify the critical distance, the experimental endurance limit is employed for the case where $\sigma_{\infty\text{ref}}=85$ MPa, and $\sqrt{\text{area}}_{\text{ref}}=400$ $\mu$m (Section \ref{sec:ExpSum}). This results in a value of 79 $\mu$m for $L/2$.
\subsection{Gradient criterion}
Based on the initial work of Nadot \cite{Nadot.06}, Gadouini et al \cite{Gadouini.08} described the impact of a defect on fatigue resilience by amplifying the equivalent stress by a function of the surrounding stress gradient. This criterion is expressed as:
\begin{equation}
\sigma_{\text{eq}}^\ast=\sigma_{\text{eq}_{\text{Vu},\text{Max}}}\left(1-b_g\frac{\triangledown\sigma_{\text{eq}_{\text{Vu}}}}{\sigma_{\text{eq}_{\text{Vu},\text{Max}}}}\right) \leq \beta
\end{equation}
with the equivalent stress gradient expressed as:
\begin{equation}
\triangledown\sigma_{\text{eq}_{\text{Vu}}}=\frac{\sigma_{\text{eq}_{\text{Vu},\text{Max}}}-\sigma_{\text{eq}_{\text{Vu},{\infty}}}}{\sqrt{\text{area}}}
\end{equation}
where $\sigma_{\text{eq}_{\text{Vu},\text{Max}}}$ is the maximum equivalent stress calculated at the tip of the defect and $\sigma_{\text{eq}_{\text{Vu},{\infty}}}$ is the nominal equivalent far-field stress. Having the same dimension as $\sqrt{\text{area}}$,  the length parameter $b_g$ requires evaluation before this criterion can be applied. This parameter was identified with the same experimental data used for the CDM approach (Section \ref{sec:critdistance}), with $\sigma_{\infty\text{ref}}=85$ MPa and $\sqrt{\text{area}}_{\text{ref}}=400$ $\mu$m such that:
\begin{equation}
b_g=\sqrt{\text{area}}_{\text{ref}}\left(\frac{\sigma_{\text{eq}_{\text{Vu},\text{Max}}}-\beta}{\sigma_{\text{eq}_{\text{Vu},\text{Max}}}-\sigma_{\text{eq}_{\text{Vu},{\infty}}}}\right)
\end{equation}
\section{Comparison between simulations and experimental results}\label{sec:comparison}
\begin{figure}
\centering
      \includegraphics[width=4 in]{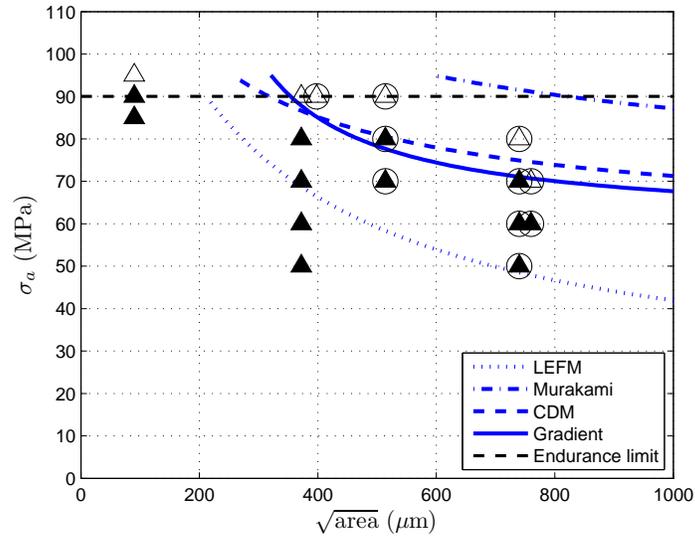}
      \caption[]{Comparison between tensile simulations and experimental results. Circled points indicate artificial defects.\label{fig:TenComp}}
\end{figure}

\begin{figure}
\centering
      \includegraphics[width=4 in]{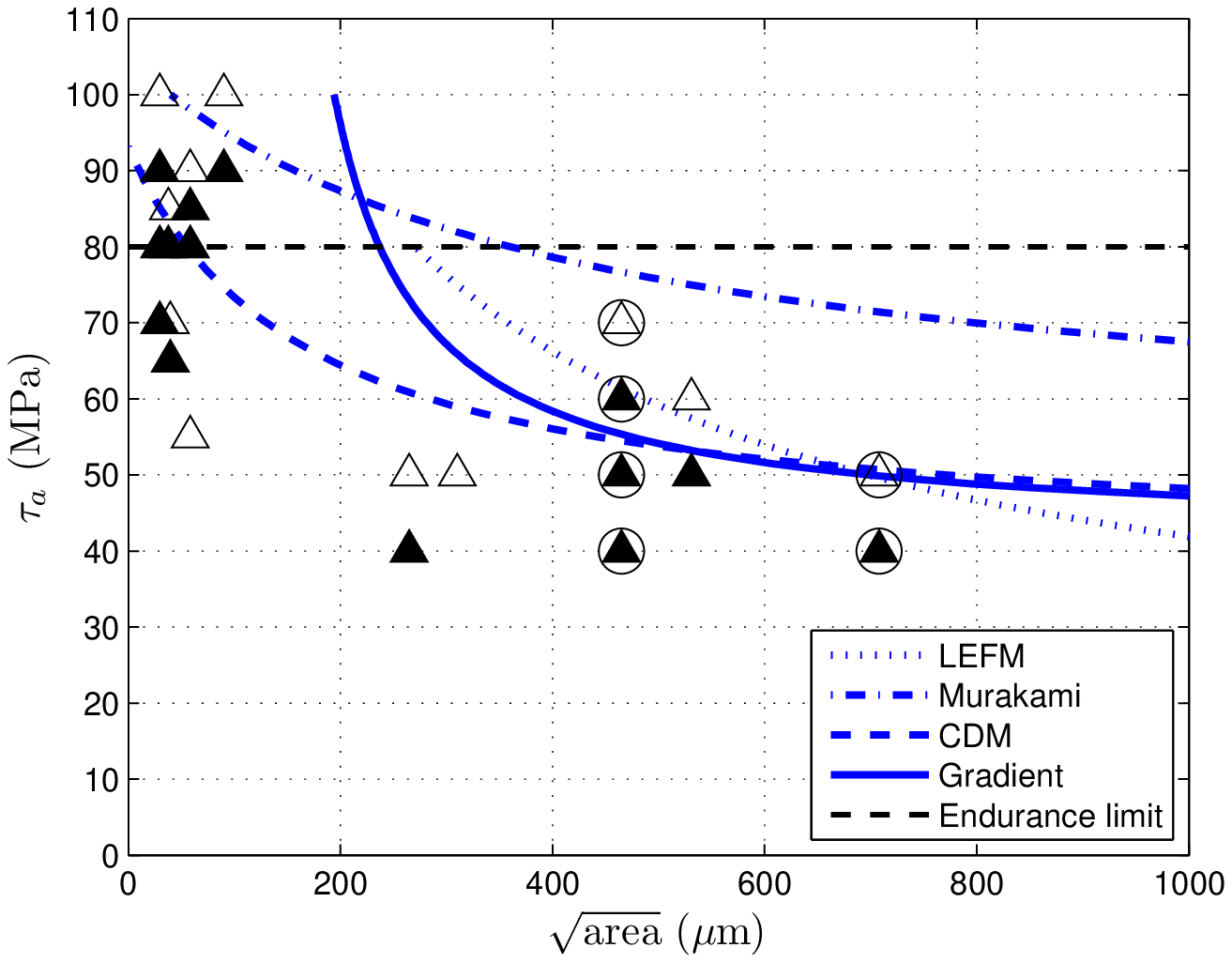}
      \caption[]{Comparison between torsion simulations and experimental results. Circled points indicate artificial defects.\label{fig:TorComp}}
\end{figure}

\begin{figure}
\centering
      \includegraphics[width=4 in]{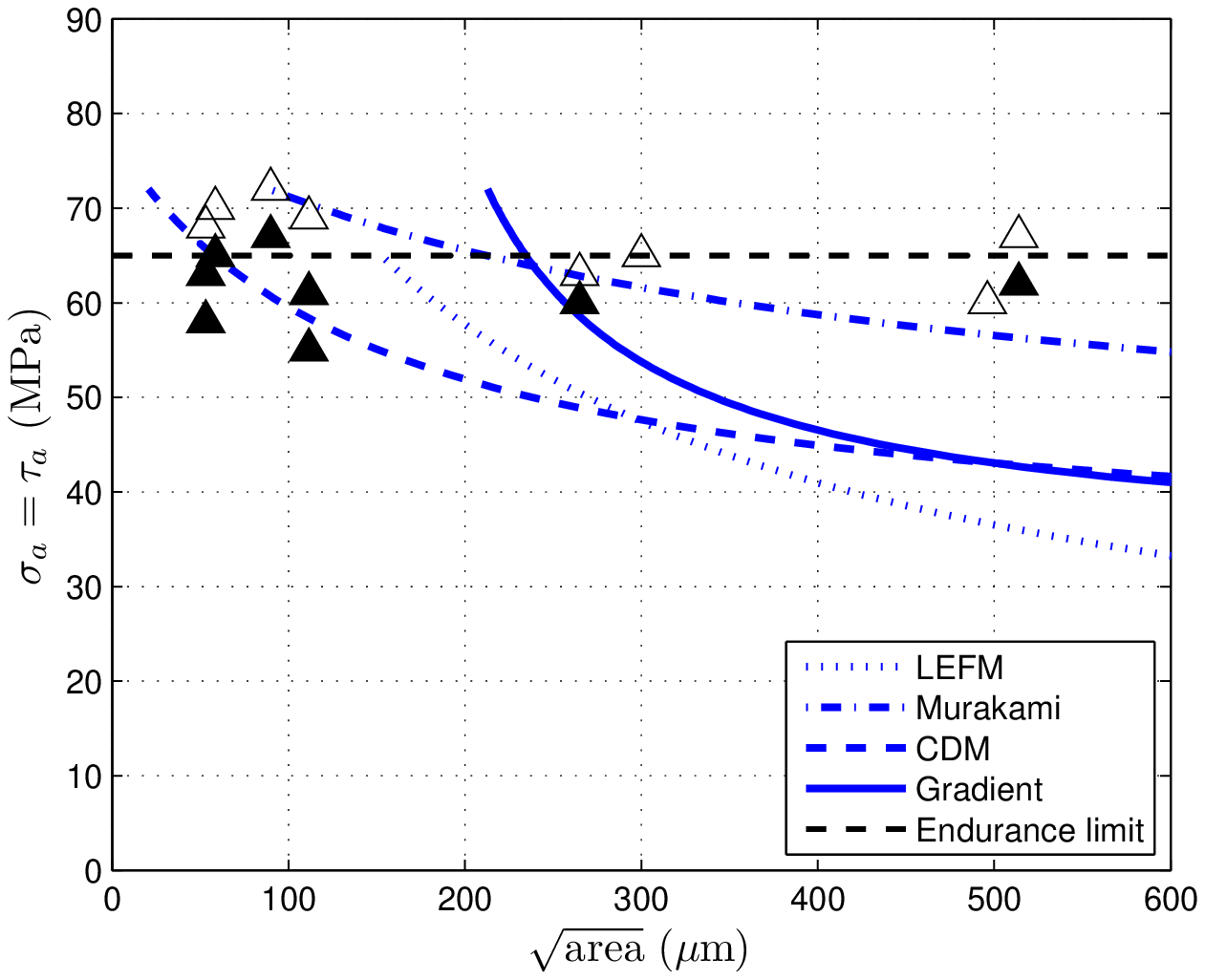}
      \caption[]{Comparison between tension-torsion simulations and experimental results\label{fig:TenTorComp}}
\end{figure}

Figures \ref{fig:TenComp}, \ref{fig:TorComp} and \ref{fig:TenTorComp} present the predictions made by each of the four approaches compared with experimental results for the three loading cases. Under tension (\ref{fig:TenComp}), the CDM and Gradient approaches describe the experimental results quite well. However, this is likely due to the single experimental point employed for identification of the necessary parameters. The trend of the fatigue limit versus defect size is also well described. Murakami's equation leads to non-conservative results, but the trend is well described. LEFM gives conservative results, with an acceptable endurance limit trend. Under torsion (\ref{fig:TorComp}), all approaches adequately describe both the fatigue limit and trend; however, Murakami relationship is again non-conservative and the CDM approach leads to a small allowable defect size. Under combined loading (\ref{fig:TenTorComp}), all approaches are non-conservative except Murakami. Again, the critical defect size is largely underestimated by the CDM.

\begin{figure}
\centering
      \includegraphics[width=4 in]{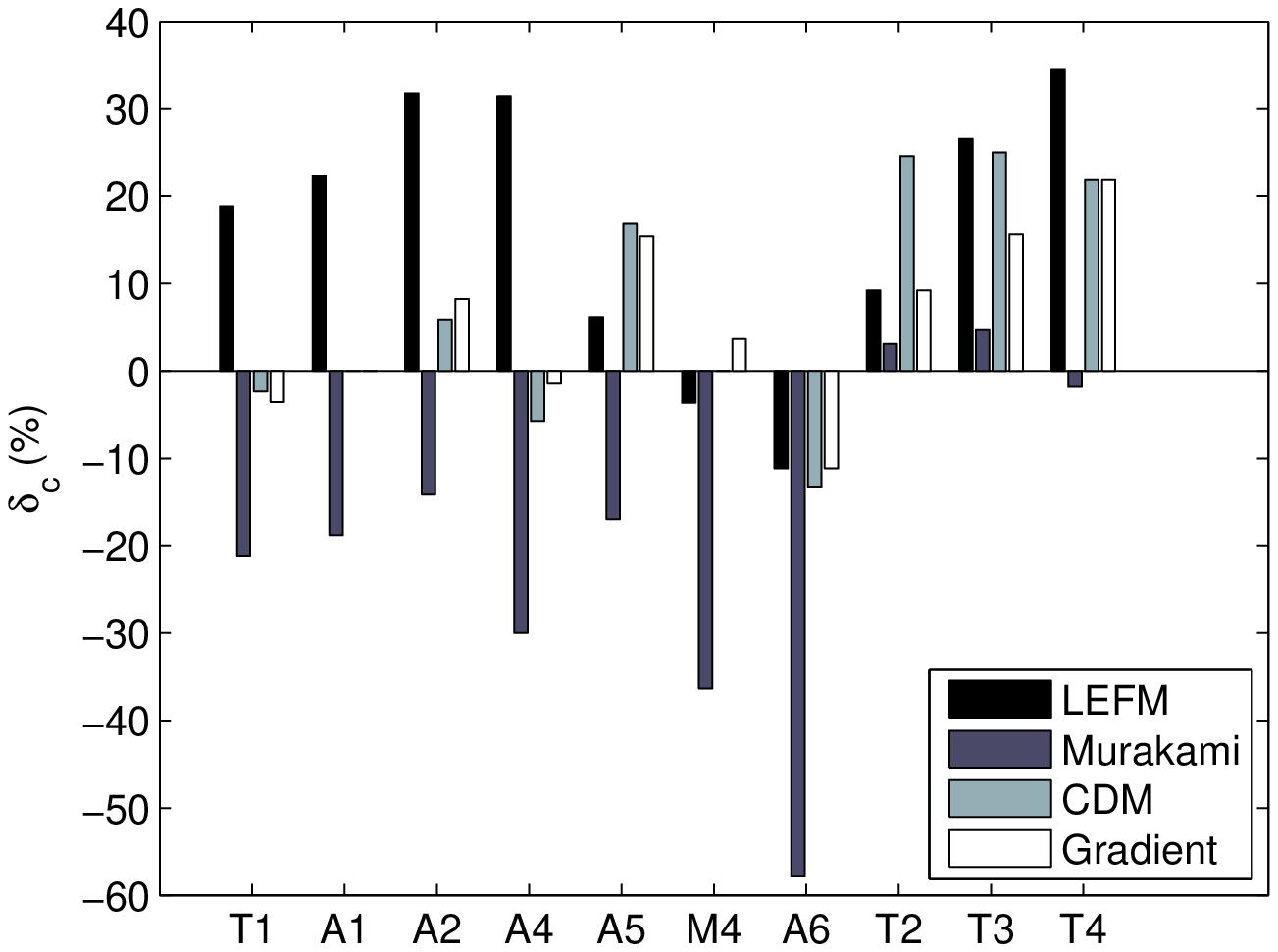}
      \caption[]{Quantitative comparison of experimental results for each of the four simulation approaches. Specimens A1, A2, A4 and A5 contained artificial defects.\label{fig:QualComp}}
\end{figure}

Table \ref{table:comparison} demonstrates the error between experimental and simulated endurance limit results for selected conditions. The difference between the calculated and experimentally found endurance limits ($\delta_c$) is presented in Figure \ref{fig:QualComp}. Non-conservative assessments are identified by negative values and positive values indicate conservative ones. The result obtained by averaging the absolute error given by each approach leads to the following result: LEFM = 19 \%, Murakami = 20 \%, CDM = 11 \% and Gradient = 9 \%. Based on this comparative assessment, describing the effect of a defect through the elastic stress field (Gradient or CDM approaches) provides better results than the LEFM or Murakami methods. However, Murakami's equation remains very good at estimating the endurance limit of A356 for different defect sizes and loading cases using only the macroscopic hardness of the material. LEFM provides conservative estimates with approximately the same average error for the same identification cost: the only experimental parameter required is the effective threshold stress intensity factor for long cracks. Both the CDM and Gradient approaches are more accurate, however they require the determination of the elastic stress field around the inclusion as well as three experimental endurance limits, including one with a defect. It is important to note that computation of elastic stresses is quite important in these latter approaches for A356-T6 as the endurance limit is half the yield stress. This indicates that there is a very small amount of plasticity at the tip of the defect.
\begin{table}[h!]
\centering
\caption{Comparison between experimental and calculated endurance limits. Specimens A1, A2, A4 and A5 contained artificial defects.\label{table:comparison}}
    \begin{singlespacing}
    \begin{tabular}{lccccc}
    \hline
    \multirow{2}{*}{Specimen} & \multirow{2}{*}{Loading}& \multicolumn{4}{c}{$\delta_c$ (\%)}\\
    &	&	LEFM & Murakami & CDM & Gradient\\
    \hline
T6 &\multirow{4}{*}{Tension}& 18.8 &-21.9 & -2.35 & -3.53\\
A1$^{\text{a}}$ &	& 22.4 & -18.8 & 0 & 0\\
A2&	& 31.8 & -14.1 & 5.88 & 8.24\\
A4&	& 31.4 & -30.0 & -5.71 & -1.43\\
\hline
A5&\multirow{3}{*}{Torsion}& 6.15 & -16.9 & 16.9 & 15.4\\
M4&	& -3.64 & -36.4 & 0 & 3.64\\
A6&	& -11.1 & -57.8 & -13.3 & -11.1\\
\hline
 T2&\multirow{3}{*}{Combined}&9.23 & 3.08 & 24.6 & 9.23\\
T3&	& 26.6 & 4.69 & 25.0 & 15.6\\
T4& & 34.5 & -1.81 & 21.8 & 21.8\\
\hline
\end{tabular}
\end{singlespacing}
\newline
$^{\text{a}}$\small{$\sqrt{\text{area}}_{\text{ref}}$} condition\\
\end{table}
\section{Conclusion}
\begin{itemize}
\item For A356-T6 submitted to multiaxial fatigue loading, fatigue cracks can initiate either on casting defects or inside the microstructure. Both scales are in competition for the localization of cyclic plastic deformation that leads to the initiation of a crack.
\item When a crack initiates at a defect, the various types of defect can be characterized by $\sqrt{\text{area}}$: natural defects were found to have the same endurance limit as artificial defects.
\item The critical defect size has been found to be 400 $\pm 100$ $\mu$m in A356-T6. This result was obtained for both artificial and natural defects under three fully reversed ($R=-1$) loading scenarios: tension, torsion and combined loading.
\item Further experimental effort is needed to better characterize the Kitigawa relationship in pure torsion, understand the effect of small defects on all loading scenarios and to study the impact of larger defects in combined tension-torsion loading.
\item Multiaxial Kitagawa-type relationships were simulated using four different approaches: Murakami, LEFM, CDM and the Gradient method. Results show that Murakami relationships give mainly non-conservative results with an average error of 20\%. LEFM provides mainly conservative estimates with an average error of 19\%. The error in using the CDM and Gradient approaches are both equivalent with mainly conservative results and an average error of  11 and 9\%, respectively.
\end{itemize}
\clearpage
\end{document}